%% file: arxiv.tex
\documentclass[sigconf]{acmart}
\input{meta/packages}

\AtBeginDocument{%
  \providecommand\BibTeX{{%
    \normalfont B\kern-0.5em{\scshape i\kern-0.25em b}\kern-0.8em\TeX}}}

\acmConference[arXiv Pre-print]{arXiv Pre-print}{}{}
\acmBooktitle{arXiv Pre-print}
\renewcommand\footnotetextcopyrightpermission[1]{}
\makeatletter
\renewcommand\@formatdoi[1]{\ignorespaces}
\makeatother

\begin{document}
\fancyfoot{}
\fancyhead{}
\input{meta/title}
\pagestyle{plain}
\input{meta/authors}

\begin{abstract}
\input{00-abstract}
\end{abstract}

\input{meta/ccs}

\maketitle
\input{00-all}

\begin{acks}
\input{99-acknowledgements}
\end{acks}

\bibliographystyle{format/ACM-Reference-Format}
\balance
\bibliography{references}

\end{document}

%% file: meta/packages.tex
\usepackage{enumitem}
\setlist{listparindent=1.25em}

\usepackage[figuresright]{rotating}
\usepackage{tabularx}
\usepackage{balance}

\newcommand{\showcase}{showcase.codeday.org/projects/labs}

%% file: meta/title.tex
\title{Open-Source Internships With Industry Mentors}

%% file: meta/authors.tex
\author{Tyler Menezes}
\email{tylermenezes@codeday.org}
\orcid{0000-0002-7975-2533}
\affiliation{
 \institution{CodeDay}
 \city{Seattle}
 \state{Washington}
 \country{USA}
}

\author{Alexander Parra}
\email{alexparra@codeday.org}
\affiliation{
 \institution{CodeDay}
 \city{Seattle}
 \state{Washington}
 \country{USA}
}
\orcid{0000-0001-7526-5765}

\author{Mingjie Jiang}
\email{mingjie@codeday.org}
\orcid{0000-0001-7064-7019}
\affiliation{
 \institution{CodeDay}
 \city{Seattle}
 \state{Washington}
 \country{USA}
}

%% file: 00-abstract.tex
Internships help students connect what they have learned in the classroom to the real world, and students with access to internships are more likely to graduate and secure employment. However, many students are unable to find an internship by the time they graduate.

This experience report describes a program where volunteer software engineers mentor students as they work on open-source projects in the summer, offered as an alternative to a traditional internship experience. We catalog the considerations involved in providing an experience similar to a traditional internship, describe our program's design, and provide two years' worth of participant evaluations and career outcomes as a measure of efficacy.

The program served mostly undergraduates from non-R1 schools who are underrepresented in technology, and achieved similar educational outcomes to a traditional internship program. Most promisingly, mentors were willing to serve as a professional reference for 80\% of students and the number of graduating seniors who secured full-time employment in technology was 7 points higher than average (despite occurring during the COVID-19 pandemic).

%% file: meta/ccs.tex
\ccsdesc[500]{Social and professional topics~Informal education}
\ccsdesc[500]{Social and professional topics~Software engineering education}

\keywords{open source software, internships, industry engagement}

%% file: 00-all.tex
\input{01-introduction}
\input{02-background}
\input{03-model}
\input{04-results}
\input{05-future-work}
\input{06-conclusions}

%% file: 01-introduction.tex
\section{introduction}

Computer science coursework provides students with the fundamental skills to learn what they need to know later in their careers, but many colleges engage industry to show students how to apply what they learn \cite{beaubouefComputerScienceCurriculum2011}. Industry engagement can come in many forms---including mentoring, guest speakers, tours, case studies, and capstone projects---but one of the most significant predictors of post-graduation employment is participation in an internship \cite{callananAssessingRoleInternships2004}.

In this report, we describe an open-source internship, which we define as an industry engagement experience in which students work on a new or existing software project, licensed under an open-source license, with the guidance of an experienced software engineer. Many prior studies and initiatives have shown that these open-source projects can provide a real-world environment for learning and improve diversity \cite{postnerSurveyInstructorsExperiences2018,hislopOpenSourceExtracurricular2019,hislopStudentReflectionsLearning2020,ellisPowerOpenSource2021}.

Compared to internships where students work on proprietary software, most open-source internships do not require access to corporate resources. While some programs have explored paid mentors \cite{silvaStudentsEngagementOpen2017}, our program used volunteer mentors exclusively.

Our goal was to evaluate the potential of open-source internships to (a) provide similar outcomes to a traditional internship, and (b) to reach students who were otherwise unlikely to have one.

%% file: 02-background.tex
\section{Internships and CS Education}

Most undergraduates pursuing a computer science degree chose the major because they believe it will improve their job prospects in industry \cite{nortonPerceivedBenefitsUndergraduate2017,alshahraniUsingSocialCognitive2018,helpsStudentExpectationsComputing2005}, but a disconnect between coursework and industry has long been reported by both graduates \cite{begelStrugglesNewCollege2008,craigListeningEarlyCareer2018,kapoorUnderstandingCSUndergraduate2019} and employers \cite{begelStrugglesNewCollege2008}. Institutions commonly try to resolve this disconnect through industry engagement opportunities such as capstone projects, mentoring, guest speakers, and internships.

Internships in particular provide benefits that are difficult to replicate in the classroom:

\begin{itemize}
    \item \textbf{Functioning on a team:}
    Software engineers must learn to deal with resistance from co-workers or managers or delays from others on tasks that block their progress and must likewise learn to prioritize and communicate their work to co-workers and managers \cite{beaubouefComputerScienceCurriculum2011}.
    
    \item \textbf{Career confidence:}
    Students who believe they have a path into a career put more time into educational activities and are more likely to overcome obstacles. The real-world practice, goal-setting, and performance feedback afforded by an internship can increase this confidence \cite{lentUnifyingSocialCognitive1994}.

    \item \textbf{Recruiting and retention in the major:}
    Few students enroll in STEM majors, and many drop out \cite{chenSTEMAttritionCollege2013}. Studies have shown that the number and diversity of students entering and staying in these majors can be increased by providing internships \cite{frylingCatchEmEarly2018} and research experience \cite{dahlbergImprovingRetentionGraduate2008, tashakkoriEarlyParticipationCS2011}.

    \item \textbf{Securing a Job After Graduation:}
    Studies have found that whether a student had completed an internship or not is one of the most significant variables as to whether or not they have a job after graduation \cite{callananAssessingRoleInternships2004, jonesTransformingCurriculumPreparing2002, knouseRelationCollegeInternships1999, saltikoffPositiveImplicationsInternships2017}, their starting salary, and the amount of time they spend looking for a job \cite{gaultUndergraduateBusinessInternships2000}. This relationship holds even for unpaid interns \cite{saltikoffPositiveImplicationsInternships2017}.
\end{itemize}

\subsection{Access to Internships}
For many students, internships are hard to come by. Among students seeking bachelor’s degrees, only 20\% of rising sophomores get internships, and less than half of rising juniors/seniors do. Even by graduation, only 60\% of students have internship experience \cite{kapoorExploringParticipationCS2020, kocClass2014Student2014}.

Access to internships is not equitable: students with a high household income are much more likely to get internships. One study found that, while most students with a household income over \$150,000 per year were offered an internship by the time they graduated, the rate dropped to only 35\% for students with a household income under \$100,000 per year \cite{kapoorExploringParticipationCS2020}. Although further studies are needed, it's likely that gender and race are additional factors, given the technology industry's reputation for being unwelcoming. The name recognition of the institution at which a student studies likely has a significant effect as well: indeed, implementation of this program was driven by the experiences of students attending small affordable schools, who expressed frustration that their schools were ignored by recruiters.

%% file: 03-model.tex
\section{CodeDay Labs Internship Model}

An open-source internship is a form of industry engagement which consists of at least the following three features:

\begin{itemize}
    \item Students create a new software project or contribute features or bug-fixes to an existing software project.
    \item Someone with current or past work experience as a full-time software engineer provides mentorship and supervision to the student.
    \item The resulting work is licensed under an Open Source Initiative approved license \cite{OpenSourceDefinition}.
\end{itemize}

Our open-source internship program, CodeDay Labs, recruited volunteer software engineers to mentor, with each mentor responsible for choosing a project to work on.

A team of 2--3 students were assigned to work together under the guidance of a mentor for 6 to 12 weeks. (The length was dependent on whether the student could use the experience for school credit if it met an hours requirement.) Students collaborated using synchronous calls, as well as asynchronously using a chat program, issue tracking software, and online code reviews.

Mentors supervised and guided students in twice-weekly, hour-long check-ins, in at least two fifteen-minute one-on-one meetings with each student, and through written performance evaluations every 1-2 weeks.

To reduce the time commitment required from mentors (and thus increase the number of mentors who can help), we hired TAs to provide detailed debugging help. Students could schedule time with these TAs through an online portal.

\subsection{Addressing Barriers and Equity}

A goal of this program was to increase equal access to an internship experience, but providing more internship opportunities would not do this alone. We identified three factors that contribute to students’ inability to secure a traditional internship and designed the program and application process to address them:

\begin{enumerate}
    \item \textbf{Lack of preparation:}
    Many students do not adequately prepare to get internships. One study found that 37\% of students who did not receive an internship attributed it to failing to take the actions needed \cite{kapoorBarriersSecuringIndustry2020}. Students who secure internships spend a median of 3 hours per week writing applications, attending career fairs, working on personal projects, and practicing interviews, compared to a median of 1 hour for those who did not \cite{kapoorExploringParticipationCS2020}. Although GPA does not correlate with success in finding an internship, \cite{gaultUndergraduateBusinessInternships2000} 23\% of students in one study relied on GPA alone when applying for internships. \cite{kapoorExploringParticipationCS2020}

    With this in mind, we designed the application with several options to allow students to demonstrate their proficiency in a way they are likely already prepared for. Specifically, reviewers considered: resume/LinkedIn, classes taken, participation in hackathons and other CS events, passion projects, a personal essay, and technical interview questions. Each aspect was optional, and we encouraged students to provide the evidence they already had.

    \item \textbf{Competing priorities:}
    A 2020 study found that 34\% of students who did not secure an internship said it was because the time to secure an internship or the time involved in carrying on an internship would conflict with a higher life priority. \cite{kapoorBarriersSecuringIndustry2020}
    
    Our preliminary study of students indicated common barriers were: coursework and maintaining a good GPA, being a caretaker for a family member, health concerns, or being unable to quit a job they rely on for year-round financial support. To support these students, we allowed them to choose a time commitment. Initially, we provided three options – 10, 20, or 30 hours a week – but removed the 10-hour option in the second year because it was unpopular.

    \item \textbf{Low self-efficacy:}
    In one study of 300 students who had not received an internship, nearly half had self-selected out of applying because of a lack of confidence: because they felt their academic standing was too low, because they felt their resume would not stand out, or because they otherwise thought they lacked the experience to succeed in an internship. \cite{kapoorBarriersSecuringIndustry2020}
    
    Social cognitive career theory suggests that students are more likely to be interested when they believe they have a chance at acceptance. \cite{lentUnifyingSocialCognitive1994} Accordingly, we partnered with professors and colleges to personally inform students they were a good fit. Once students opened the application, we offered email, phone, and live-chat support to encourage students to ask questions about eligibility and selection.
\end{enumerate}

As most interns in the technology industry are paid, no discussion of equity would be complete without addressing payment. Partners funded stipends for 39 students (ranging from \$2,000 to \$4,000), but the rest of the students were not paid. We believe that students generating business value should be paid for their work, which motivated this program's exclusive focus on contributions to open-source projects with no business value. We believe the educational benefit and lack of business value justified admitting more students than we could provide stipends. Nonetheless, we are seeking industry partners to fund more stipends in the future.

\subsection{Project and Mentor Selection}

Mentors were recruited using LinkedIn posts and advertisements, by reaching out to software engineering leaders, and through partnerships with other programs. Each mentor filled out an application to screen their background, and then a member of program staff held an individual 15-minute phone call to finalize details.

As part of the application, mentors proposed projects that were one of two types: improvements to an existing open source project or creation of a new open source project. Because we desired that projects teach the skills needed to succeed in the workforce, we conducted a survey of hiring managers and created a list of requirements which each project would need to meet. (Table~\ref{tab:competencies}) Program staff worked with mentors to conform projects to these requirements. Two examples of projects are presented below. A full list of projects is available on the web at \showcase.

\input{tab-competencies}

\subsubsection{Example of New Open-Source Project}
\begin{quote}
LiDAR is a way of obtaining an accurate 3D representation of a scene, often used in self driving cars to detect their surroundings.

Most analysis done on data like this is done without visualizing all the data together, because it's in different formats/too large to efficiently visualize in real time. This project would involve taking this data, converting it to a standard format that Cesium's pipeline can ingest (likely using Python since it has many of the helper libraries you'll need), and then building an application to visualize it in 3D using Cesium's JavaScript library.
\end{quote}

\subsubsection{Example of Improving an Existing Open-Source Project}
\begin{quote}
Crates.io is the default, public package registry used by rust developers everywhere. Developers often want the means to privately publish crates (rust packages), so they can continue to follow best practices to version and release software internal to their teams or businesses.

Current solutions for a private crates registry are hard to find and very costly. However, there exists an open source implementation of the crates registry API one can easily run on their local machine - "Alexandrie".

In this internship, we will build on Alexandrie to provide an open-source solution that others can use to more easily deploy a private crates registry to cloud providers. We will be using docker, and developing the reference solution to be deployable to a Kubernetes cluster in one of the major cloud providers.
\end{quote}

\subsection{Student Selection}

To select students for the program, the admissions team (comprising mentors, program staff, and alumni) read each application\footnote{Some students who applied for the program were recommended by faculty at a partner college. These "direct admits" did not go through this selection process. We did not observe any differences between these students and those accepted through the open application.} and first evaluated whether the applicant met the technical bar:

\begin{itemize}
    \item Demonstrated passion for CS (e.g., by taking classes, joining clubs, working on projects, attending events, or a personal statement)
    \item Demonstrated knowledge in algorithms and data structures
    \item Ability to read and understand code written by others
    \item Ability to do independent research to solve a problem
    \item Knowledge of collaboration tools such as Git
\end{itemize}

Secondly, the admissions team evaluated the student’s access to internship opportunities. Students who were closer to graduating with limited experience were favored.

Application were scored from 1-5, and the admissions software continuously ranked applicants based on the number of scores and a margin of error to account for differing numbers of scores. Admissions were offered to the top-ranked students as space became available, ending two weeks before the program.

\subsection{Matching}
Students matched with a mentor/project in two phases.

First, we used the Elastic search engine to produce a ranked list of 25 recommendations for each student, using technology proficiency, interest, and timezone information from student/mentor applications. Students chose their top six projects from this list and ranked them in order of preference. The system also evaluated the popularity of projects: as more students selected projects, they became less likely to appear in others’ recommendations.

Second, a modified Gale-Shapely algorithm \cite{galeCollegeAdmissionsStability1962} was used to match students to projects. Students and mentors received an email introduction several days before the start of the program.

\subsection{Additional Features of Internships}

Although all internships involve supervised experiential learning guided by industry, many leading companies provide other opportunities for their interns, \cite{cunninghamBuildingPremierInternship2012} which we attempted to replicate:

\begin{itemize}
    \item \textbf{Interview experience:} The process of applying for an internship helps students practice the same skills necessary to secure full-time employment later.
    
    Although we did not conduct individual interviews as part of admissions, students were able to use a web portal to request interview feedback and practice interviews with software engineers, hiring managers, and HR employees.

    \item \textbf{Professional networking and advice:} Many successful internships provide students a chance to meet both near-peers and leaders, allowing students to gain career advice beyond what their school can offer.
    
    Students in our program were able to connect with their mentors in dedicated 1-1 sessions, and we also hosted 2-5 career panels each week during the program.
    
    \item \textbf{Technical training:} Some companies offer in-house training or the ability to attend technical talks, in order to increase the breadth of industry-connected skills the interns know.
    
    During the program we hosted 1-3 technical talks each weekday.
    
    \item \textbf{Final presentation:} Presentations help interns develop skills and gain confidence in technical presentations.
    
    At the end of the program, students created "tech talks": 10-15 minute presentations describing the technology used, to be shared with potential employers. Some mentors also invited students to present to co-workers.
\end{itemize}

%% file: tab-competencies.tex
\begin{sidewaystable*}
\vspace*{45\baselineskip}
\caption{Project Competencies}
\label{tab:competencies}
\begin{tabularx}{\linewidth}{|l|X|X|}
\hline
 && \\
 & \textbf{Core Competencies\footnote{Projects included all core competencies}} & \textbf{Advanced Competencies\footnote{Projects included 2 or more advanced competencies, at least one of which was technical. Projects with 4+ were recommended to more advanced students.}} \\
  && \\ \hline && \\
\textbf{1. Software} & A.                 Identifying and defining problems using debugging techniques. &  \\
\textbf{Engineering Process} & B.                  Online and peer research to discover existing solutions to a problem. &  \\
 & C.                  Experimentation; learning by doing. &  \\
 & D.                 Developing and evaluating a set of proposed solutions to a problem. &  \\
 & E.                  Verifying that a problem is solved. &  \\
 & F.                  Documenting a solution for others. &  \\
  && \\ \hline && \\
\textbf{2. Interpersonal} & A.                 Working collaboratively and productively in a team. & \textbullet\ Technical writing. \\
 & B.                  Individual task management in an agile workflow. &  \\
 & C.                  Managing change and uncertainty. &  \\
 && \\ \hline && \\
\textbf{3. Cross-Functional} & A.                 Requirements gathering. & \textbullet\ Systematic thinking and architecture design. \\
 & B.                  Technical speaking / presentations. & \textbullet\ Project management. \\
 &  & \textbullet\ Speaking with customers and incorporating feedback. \\
 &  & \textbullet\ Risk management. \\
 &  & \textbullet\ User interface design. \\
 &  & \textbullet\ Business needs analysis and/or business case justification. \\
  && \\ \hline && \\
\textbf{4. Technical} & A.                 Software and/or hardware architecture. & \textbullet\ User analytics and data-driven design \\
 & B.                  OOP and/or functional programming. & \textbullet\ Statistics and data analysis. \\
 & C.                  Testing and quality assurance methodologies. & \textbullet\ Discrete mathematics. \\
 & D.                 Creating/refactoring and documenting code in a reusable manner. & \textbullet\ Machine learning. \\
 & E.                  Setting up and using modern development environments. & \textbullet\ API architectures, tradeoffs, and design. \\
 &  & \textbullet\ Consuming APIs. \\
 &  & \textbullet\ Cloud deployment and/or system administration. \\
 &  & \textbullet\ Containers and/or orchestration. (e.g. Docker, Kubernetes) \\
 &  & \textbullet\ Event programming methodologies (e.g. Kafka) \\
 &  & \textbullet\ Evaluating and improving system performance. \\
 &  & \textbullet\ Algorithm design and development. \\
 &  & \textbullet\ Distributed systems. \\
 &  & \textbullet\ Data modeling. \\
 &  & \textbullet\ Database design and development. \\
  && \\ \hline
\end{tabularx}
\end{sidewaystable*}

%% file: 04-results.tex
\input{fig-checkins}

\section{Results and Evaluation}

We operated the program in the summers of 2020 and 2021. Except as noted previously, the program was substantially the same in both years. In total 311 students were admitted (197 in 2020 and 114 in 2021).\footnote{The full program also included a "beginner track" to which an additional 145 students were admitted. This track was for high school students who had only completed AP CS A, mentored by college students rather than industry professionals, and students in this track would not qualify for traditional internships. All data from the beginner track was omitted.} All but 5 students completed the program.

80\% of students were undergraduates; specifically: 29\% seniors, 33\% juniors, 27\% sophomores, and 10\% freshmen. 97\% of students had never completed an internship before.

Only 19\% of undergraduates who participated attended an R1 university (which are the schools most commonly targeted by recruiters from technology companies). 30\% attended a non-R1 Doctorate-/Masters granting university, 43\% a Baccalaureate granting college, and 8\% an Associates granting college. In total students came from 77 unique schools.

67\% of students were under-represented in technology (51\% were women or non-binary, and 31\% were Black, Latinx, or Native American). Although the program was conducted virtually, 86\% of students were located in North America with the remaining 14\% located in Europe and Asia.

Students who received a stipend were slightly more likely to commit to spending 40 hours per week on their project rather than 20, but we did not observe any differences in outcomes, so we have chosen not to report this data separately.

We matched these students with 125 mentors. Although mentors were volunteers who did not represent their employers, we collected each mentor's employment history. The most popular employer was Microsoft, which accounted for 10\% of mentors. 63\% worked for a Fortune 500 company at the time of mentoring. Mentors had a median of 3 years of full-time work experience. 62\% of mentors had a job title corresponding to Software Engineer, 20\% had a title corresponding to Senior Software Engineer, and 9\% had the title corresponding to Engineering Manager. The most senior mentor was a VP of Engineering for a 200-person technology company.

To evaluate the success of the program, we considered long-term job outcomes, mentor evaluations, and student self-evaluations.

\subsection{Job Placement Outcomes}

Ultimately, we hope internships will lead to employment, and measuring the rate of students who obtained employment was our primary evaluation criteria. Because no one source of data is perfect, to determine employment status we combined information from student follow-up surveys, school reporting data, and public information from LinkedIn and Github.

The NCES reports that one year after graduation, an average of 61.9\% of students with a bachelor's degree in Computer Science secure employment in their field. \cite{DigestEducationStatistics2021} We found that at least 69\% of the graduating seniors who participated in the 2020 program\footnote{We are not able to present longitudinal data for 2021's program due to the timeline of publication.} had secured a full-time job in technology within nine months of the end of the program, representing a 12\% or greater increase from baseline. This number may be artificially deflated by a lack of available data and suppressed new-grad hiring as a result of the COVID-19 pandemic.

Additionally, we found that at least 60\% of rising seniors had secured a traditional internship within nine months.

\subsection{Mentor Skill Evaluations}

A primary aim of the program was to prepare students for industry work. We asked mentors to evaluate how close each of their mentees was to meeting expectations for a new-grad hire, considering: their ability to work independently through the software development process, their interpersonal skills, and their technical proficiency. (We did not have mentors evaluate cross-functional competencies as these were a part of the larger program.) Mentors could choose "yes, the student already meets new-grad expectations", "on-track to meet by graduation", or "no, needs work". Mentors were asked to assess student skill at the end of each week, as well as provide a final evaluation.

Weekly results were converted to a numeric score (yes = 1, by graduation = 0.5, no = 0) and an average was graphed over time. (Figure \ref{fig:checkins}) Over the course of the program, mentors' ratings of their students improved across all categories by at least 30-40\%. Most notably, student competency in the software development process improved by 60\% between weeks 2-4. (This corresponds to student remarks from the first two weeks that they were struggling to succeed without the direct guidance they were used to from school.)

The final results, presented in Table \ref{tab:mentors}, paint a positive picture, with mentors indicating that over half of students already meet their expectations and they expect 90\% to do so by the time they graduate.

Mentors also indicated they would be willing to provide a positive employment reference for 82\% of students who participated in the program. (That number increased to 95\% for students who were rated as "on-track to meet by graduation" or higher in all four evaluation criteria.)

\input{tab-mentors}

Overall, mentors said they agreed with the decision to admit their mentees 91\% of the time, but only thought 64\% of students were matched with a project which was a good fit for their skill level.

\subsection{Self Skill Evaluations}

Because traditional internships have been shown to promote confidence and increase retention in the major, we surveyed students to learn how they viewed the experience.

We asked students to self-evaluate how close they were to meeting expectations for a new-grad hire using the same criteria as mentors. The results are presented in Table \ref{tab:students}. Students were overall much more optimistic than mentors, with 40\% of students rating themselves as more hirable than their mentor in at least one category. This difference was largely driven by sophomores and juniors, while seniors' assessments has no significant difference.

\input{tab-students}

We also conducted a follow-up survey of the 2020 participants three months after the conclusion of the program. In that study:

\begin{itemize}
    \item 83\% of respondents reported that the program increased their ability to work independently
    \item 74\% reported it increased their ability to work as a member of a team
    \item 71\% reported it increased their general understanding of the tech industry
    \item 59\% reported it increased their understanding of which classes they should take to benefit their career.
\end{itemize}

%% file: fig-checkins.tex
\begin{figure*}[ht]
\caption{Weekly Mentor Check-Ins For the Six Weeks of the Program, Average Scores}
\label{fig:checkins}
\includegraphics[width=\textwidth]{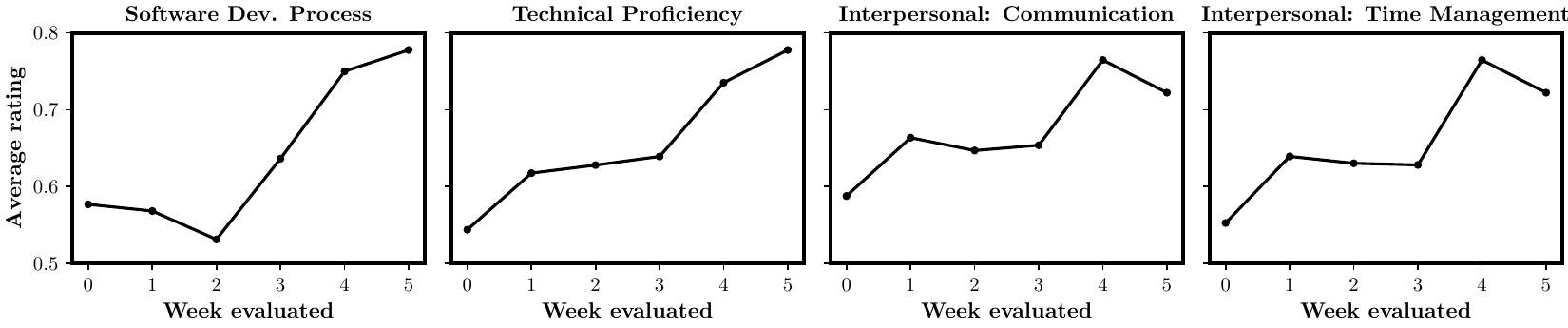}
\end{figure*}

%% file: tab-mentors.tex
\begin{table}[t]
\caption{Mentor Final Evaluation: Are Students Employable?}
\label{tab:mentors}
\begin{tabular}{llccc}
 && \textbf{Currently} & \textbf{By Graduation} & \textbf{No} \\
\multicolumn{2}{@{}l}{Engineering Process} & 56\% & 37\% & 7\% \\
\multicolumn{5}{@{}l}{Interpersonal} \\
&Communication & 56\% & 40\% & 4\% \\
&Time Management & 55\% & 40\% & 5\% \\
\multicolumn{2}{@{}l}{Technical Proficiency} & 41\% & 53\% & 6\% \\
\end{tabular}
\end{table}

%% file: tab-students.tex
\begin{table}[t]
\caption{Student Final Evaluation: Am I Employable?}
\label{tab:students}
\begin{tabular}{llccc}
 && \textbf{Currently} & \textbf{By Graduation} & \textbf{No} \\
\multicolumn{2}{l}{Engineering Process} & 87\% & 11\% & 2\% \\
\multicolumn{5}{l}{Interpersonal} \\
 & Communication & 83\% & 10\% & 7\% \\
 & Time Management & 68\% & 18\% & 14\% \\
\multicolumn{2}{l}{Technical Proficiency} & 52\% & 34\% & 14\%
\end{tabular}
\end{table}

%% file: 05-future-work.tex
\section{Future Work}

In future years we plan to grow the number of students participating in the program, which is limited by our capacity to recruit volunteers and students. We welcome collaborators from industry and open-source communities who can aid in recruiting more volunteer mentors, and colleges who are interested in providing this opportunity to their students.

Another limitation to growth is the capacity of program staff to resolve issues with individual students. Most issues involved students not managing their time correctly, and we plan to require students to commit to a specific working schedule, to include this availability in matchmaking, and to enforce the schedule with time-tracking.  The propensity of students to overestimate their skills likely also contributes to these issues, and we intend to explore ways to narrow the gap by setting realistic standards and providing increased training.

Additionally, the current matching system likely allows students too much latitude in the selection of projects, as evidenced by mentors rating 36\% of students as not a good fit for their project. Further research is needed to determine if students are simply misinformed by project descriptions, or are selecting projects which are too far outside their abilities due to ambition.

Finally, as discussed, one of the biggest drawbacks to this year's program was our inability to provide stipends to all participants. We are working to fund more stipends through private-sector partnerships.

%% file: 06-conclusions.tex
\section{Conclusions}

This experience report presented our model for an alternative internship model, provided over two years to 311 students, most of whom had no prior internships and who were less likely (demographically, and by the institutions they attended) to secure a traditional one.

Traditional internships help students connect their learning to the real-world and teach practical skills, and the results suggest that open-source internships can as well. Week-by-week mentor evaluations showed that most students entering the program were lacking in most areas needed to find employment, but that by the end they had dramatically improved in both technical proficiency --- which would likely grow as they took more classes --- and their ability to independently develop software and work with peers --- which has usually required a traditional internship.

Using follow-up data, we saw that our model also matched traditional internships in their ability to boost job placement: students who participated were at least 12\% more likely than average to find full-time employment or paid internships after the program.

Many students are unable to participate in traditional internships and, overall, our experience showed that institutions can provide these students with a similar experience using open-source software, with positive outcomes similar to traditional internships.

%% file: 99-acknowledgements.tex
This work was made possible by the work of hundreds of individuals who volunteered time and expertise. Additionally we thank we thank Cora Lewis, Erika Lamothe, and Alper Gel for finding and hosting our speakers. Thanks to Julie Cover for her work on matching. Finally, for their help in managing mentors, we thank Saharsh Yeruvas, Danielle Trinh, Maddie Willett, Sam Poder, James Click, and Hitesh Mantha.